\documentclass[%
 reprint,
superscriptaddress,
 amsmath,amssymb,
 aps,
]{revtex4-2}
\usepackage{tikz}
\usepackage{graphicx}% Include figure files
\usepackage{dcolumn}% Align table columns on decimal point
\usepackage{bm}% bold math
\usepackage{braket}
\begin{document}

\preprint{APS/123-QED}

\title{Strongest Magnetically Induced Transitions in Alkali Metal Atoms with \\ nuclear spin $3/2$}% Force line breaks with \\

\author{Armen Sargsyan}
\affiliation{Institute for Physical Research, National Academy of Sciences of Armenia, Ashtarak-2, 0203, Republic of Armenia}
\author{Ara Tonoyan}
\affiliation{Institute for Physical Research, National Academy of Sciences of Armenia, Ashtarak-2, 0203, Republic of Armenia}
\author{Rodolphe Momier}
\email[Corresponding author: ]{rodolphe.momier@u-bourgogne.fr}
\affiliation{Institute for Physical Research, National Academy of Sciences of Armenia, Ashtarak-2, 0203, Republic of Armenia}
\affiliation{Laboratoire Interdisciplinaire Carnot De Bourgogne, UMR CNRS 6303, Université Bourgogne Franche-Comté, 21000 Dijon, France}
\author{Claude Leroy}
\affiliation{Laboratoire Interdisciplinaire Carnot De Bourgogne, UMR CNRS 6303, Université Bourgogne Franche-Comté, 21000 Dijon, France}
\author{David Sarkisyan}
\affiliation{Institute for Physical Research, National Academy of Sciences of Armenia, Ashtarak-2, 0203, Republic of Armenia}

\date{\today}% It is always \today, today,
             %  but any date may be explicitly specified

\begin{abstract}
The probabilities of atomic transitions $F_e - F_g = \pm 2$ between a ground $F_g$ and an excited $F_e$ level of $D_2$ line of any alkali metal atom are zero when no external magnetic field is applied. In an external magnetic field in the range $0.1 - 3$ kG, the probabilities of these transitions called magnetically induced (MI) are highly modified. For these MI transitions, we have previously exhibited the following rule: the probabilities of MI transitions with $\Delta F = +2$ are maximal when using $\sigma^+$-polarized laser radiation, while the probabilities of MI transitions with $\Delta F = -2$ are maximal when using $\sigma^-$-polarized laser radiation. This difference has been termed Type 1 Magnetically Induced Circular Dichroism (MCD1). It is demonstrated for the first time that for alkali atoms with a nuclear spin $I=3/2$ ($^{87}\text{Rb}$, $^{39}\text{K}$,$^{23}\text{Na}$, $^7\text{Li}$) in magnetic fields $> 100$ G, the probability of the strongest $\sigma^+$ MI transition of the group $F_g = 1 \rightarrow F_e = 3'$ (transition $\ket{1,-1}\rightarrow\ket{3',0'}$) is about 4 times higher than the probabilities of the strongest MI $\sigma^-$-transitions $\ket{1,-1}\rightarrow\ket{3',-2'}$ and $\ket{2,+1}\rightarrow \ket{0',0'}$. These properties make the $\sigma^+$ MI transition $\ket{1,-1}\rightarrow\ket{3',0'}$ an interesting candidate for the study of magneto-optical processes in strong magnetic fields. 
\end{abstract}

%\keywords{Suggested keywords}%Use showkeys class option if keyword
                              %display desired
\maketitle

\section{Introduction}

According to the selection rules, the probability of so-called forbidden atomic transitions $F_e-F_g = \Delta F = \pm 2$ ($F$ is the total atomic momentum) between a ground hyperfine state $F_g$ and an excited hyperfine state $F_e$ of the $D_2$ line of alkali metals are zero when no magnetic field is applied. However, at $B > 100$ G, there is a giant increase in their probabilities. These transitions are therefore called magnetically induced (MI) \cite{Tremblay1990,Sargsyan2014LPL,Scotto2015,Scotto2016,Tonoyan2018EPL,Sargsyan2020JPB,Sargsyan2021JETPlett,Sargsyan2021PLA}. Interest in MI transitions is due to the fact that in certain intervals of magnetic fields, their probabilities can significantly exceed the probabilities of ordinary atomic transitions allowed at $B = 0$. In addition, the frequency shifts of MI transitions can reach $20-30$ GHz, which is of practical interest for the development of new frequency ranges, in particular, to stabilize the frequency of lasers on strongly shifted resonances \cite{Mathew2018}. Alkali metal atoms have about 100 MI transitions, which can be conveniently divided into two types: type-1 (MI1) and type-2 (MI2) as follows. In the basis $\ket{F,m_F}$, where $m_F$ is the projection of the total atomic momentum $F$, the first type MI1 includes transitions between the lower $F_g$ and upper $F_e$ levels $\ket{F_g,0}\rightarrow \ket{F_e = F_g, 0'}$ (the prime is used to denote the upper levels) exhibiting a zero-probability at $B=0$. However, probability grows with $B$, and with a further increase of the magnetic field $B \gg B_0$ (where $B_0 =A_{\text{hfs}}/\mu_B$ with $A_{\text{hfs}}$ the magnetic dipole interaction constant and $\mu_B$ the Bohr magneton) the probabilities asymptotically approach a constant value \cite{Olsen2011,Zentile2015,Keaveney2018}.

The second type MI2 includes transitions $\ket{F_g, m_F}\rightarrow\ket{F_e,m_F'}$,where $F_e =F_g \pm 2$ and $m_F'-m_F =0,\pm 1$. In this case, there is a steep rise in the probability of these transitions as $B$ increases. However, with a further increase of the magnetic field $B \gg B_0$, the probabilities of these transitions tend back to zero. Nevertheless, the $1\rightarrow 3'$ MI2 transitions of the $D_2$ line $^{87}$Rb can be detected up to $8$ kG magnetic fields. In this case, a strong frequency shift of $30$ GHz occurs relatively to the initial (zero-field) transition frequency. In \cite{Tonoyan2018EPL}, the following rule was established for the probabilities (intensities) of MI2 transitions: MI2 transitions obeying $\Delta F= +2$ are maximum when excited with $\sigma^+$-polarized laser radiation, while the probabilities of the MI2 transitions obeying $\Delta F = -2$ are maximum when excited with $\sigma^-$-polarized laser radiation. For some MI2 transitions, the difference between the intensities obtained when using $\sigma^+$ or $\sigma^-$ radiation can reach several orders of magnitude. This difference has been termed Type 1 Magnetically Induced Circular Dichroism (MCD1).

Revealing the strongest MI2 transitions which, in certain intervals of magnetic fields have probabilities that can exceed the ones of many ordinary (so-called allowed) atomic transitions, is important for their application in magneto-optical processes. It was demonstrated in \cite{Sargsyan2021JETPlett} that the strongest transition of the MI2 group $2\rightarrow 4'$ of $^{85}$Rb (nuclear spin $I=5/2$)  is the transition $\ket{2,-2}\rightarrow\ket{4',-1'}$ ($\sigma^+$-transition). Its intensity is approximately 2 times higher (in the range of magnetic fields $0.2$ kG - $2.5$ kG) than the intensity of the strongest $\sigma^-$ MI2 transition.

It was demonstrated in \cite{Sargsyan2021PLA} that the strongest transition of the MI2 group (transition $3'\rightarrow  5'$ of Cs, nuclear spin $I=7/2$) when using $\sigma^+$ radiation is the $\sigma^+$ transition $\ket{3,-3}\rightarrow\ket{5',-2'}$. Its intensity is approximately $2$ times higher (in the range of magnetic fields $0.2$ kG - $5$ kG) than the intensity of the strongest $\sigma^-$ MI2 transition. This difference has been termed Type 2 Magnetically Induced Circular Dichroism (MCD2).

In this work, it is experimentally demonstrated for the first time that at $B> 100$ G for atoms with a nuclear spin $I = 3/2$ ($^{87}$Rb, $^{39}$K, $^{23}$Na, $^{7}$Li) the probability of the strongest $\sigma^+$ MI2 transition of the group $1 \rightarrow 3'$ (transition $\ket{1,-1}\rightarrow\ket{3',0'}$  is about four times higher than the probabilities of the strongest $\sigma^-$ MI2 transitions $\ket{1,-1}\rightarrow \ket{3',-2'}$ and $\ket{2,+1}\rightarrow \ket{0',0'}$ numbered $3'$ and $2'$ in circles in the scheme presented in Fig. \ref{fig:1}. Theoretical calculations show that the intensity of the MI2 transition numbered $3'$ is only $1.1$ times stronger than that numbered $2'$. However, we made a more detailed comparison with the transition numbered $2'$, which is important because when $2 \rightarrow 0'$ transition is excited with $\sigma^-$ radiation, its intensity is $10^6$ times stronger than when $2\rightarrow 0'$ transition is excited with $\sigma^+$ radiation.

\section{Theoretical model}

To compute the interaction of an alkali vapor with an external static magnetic field, we use the model presented in \cite{Tremblay1990,Auzinsh2010}. In the basis of the unperturbed state vectors $\ket{F,m_F}$, the diagonal elements of the atomic Hamiltonian are given by
\begin{equation}
\bra{F,m_F}\mathcal{H}\ket{F,m_F} = E_0(F) - \mu_B g_F m_F B
\end{equation}
where $E_0(F)$ is the zero-field energy of the hyperfine state $F$, $\mu_B$ is the Bohr magneton (here chosen negative, the sign convention is discussed in \cite{arimondo}) and $g_F$ is the Landé factor associated to the $\ket{F,m_F}$ Zeeman substate \cite{steck87rb}. Off-diagonal terms of the Hamiltonian characterizing the coupling between substates due to the magnetic field are given by:
\begin{align}
&\bra{F-1,m_F}\mathcal{H}\ket{F,m_F} = \bra{F,m_F}\mathcal{H}\ket{F-1,m_F} \nonumber\\
&= -\frac{\mu_B B}{2}\left( \frac{[(J+I+1)^2-F^2][F^2 - (J-I)^2]}{F}\right)^{1/2}\\
&\times (g_J - g_I) \left( \frac{F^2-m_F^2}{F(2F+1)(2F-1)}\right)^{1/2}\nonumber\, ,
\end{align}

where $g_J$ and $g_I$ are respectively the total angular and nuclear Landé factors. The off-diagonal elements obey the selection rules $\Delta J= 0$, $\Delta L=0$, $\Delta F = 0,\pm1$. The magnetic field couples only states such that $\Delta m_F = 0$, therefore $\mathcal{H}$ exhibits a block-diagonal structure, each block corresponding to a different value of $m_F$. One can thus express the eigenvectors (states perturbed/mixed by the magnetic field) of $\mathcal{H}$ as linear combinations of the original state vectors $\ket{F,m_F}$ while only summing on $F'$ states having the same $m_F$:
\begin{equation}
\ket{\Psi (F_g,m_{F_g})} = \sum_{F_g'} \alpha_{F_gF_g'}(B)\ket{F_g',m_{F_g}}
\end{equation}
\begin{equation}
\ket{\Psi (F_e,m_{F_e})} = \sum_{F_e'} \alpha_{F_eF_e'}(B)\ket{F_e',m_{F_e}}\, ,
\end{equation}
where $\alpha_{F_{g,e}F_{g,e}'}$ are called magnetic-field-dependent mixing coefficients. The intensity $A_{eg}$ of a transition between two Zeeman sublevels $\ket{F,m_F}$ (they should be rigorously denoted $\ket{\Psi(F,m_F)}$) is proportional to the "modified transfer coefficients" such that 
\begin{align}
A_{eg} &\propto a^2[\ket{\Psi(F_e,m_{F_e})};\ket{\Psi(F_g,m_{F_g})};q] \nonumber\\
&= \left( \sum_{F_e',F_g'}\alpha_{F_eF_e'}a(F_e',m_{F_e};F_g',m_{F_g};q)\alpha_{F_g,F_g'}\right)^2
\end{align}
where $a(F_e,m_{F_e};F_g,m_{F_g};q)$ are the unperturbed transfer coefficients given by
\small
\begin{align}
a(&F_e,m_{F_e};F_g,m_{F_g};q) = (-1)^{1+I+J_e+F_e+F_g - m_{F_e}}\sqrt{2F_e+1}\\
&\times \sqrt{2F_g+1}\sqrt{2J_e+1}\begin{pmatrix}
F_e & 1 & F_g \\ -m_{F_e} & q & m_{F_g}\end{pmatrix} \begin{Bmatrix} F_e & 1 & F_g \\ J_g & I & J_e\end{Bmatrix}\, .
\end{align}
\normalsize
These coefficients involve an index $q$ reflecting the polarization of the laser ($q = 0,\pm 1$ for $\pi,\sigma^\pm$ polarization, respectively $\Delta m_F = 0,\pm 1$), and depend on Wigner 3j- (parentheses) and 6j- (curly brackets) symbols. The dipole moment component of a given transition $|\bra{e}d\ket{g}|^2$ is proportional to $A_{eg}$.

To compute theoretical spectra, the nanocell filled with the vapor is assimilated to a Fabry-Pérot microcavity following the procedure described in \cite{Dutier2002}. In this model, the transmitted signal is 
\begin{equation}
S_t \approx 2t_{wc}t_{cw}^2E_i \mathrm{Re}\left\lbrace I_f - r_w I_b \right\rbrace/|Q|^2
\end{equation}
where $t_{cw}$ and $t_{wc}$ are the transmission coefficients at the vapor-window and window-vapor interfaces respectively, $r_w$ is the reflection coefficients and $Q = 1-r_w^2\exp(2ikL)$ is the quality factor of the cavity. The transmitted signal depends on forward $I_f$ and backward $I_b$ integrals of the atomic response 
\begin{equation}
I_f = \frac{ik}{2\epsilon_0}\int_0^L P_0(z)\mathrm{d}z
\end{equation}
\begin{equation}
I_b = \frac{ik}{2\epsilon_0}\int_0^L P_0(z)\exp(2ikz)\mathrm{d}z
\end{equation}
that can be computed following the method described in \cite{Dutier2002}. This model was developed for a two-level system of resonant angular frequency $\omega_0$ but can be applied in our case by assimilating the atomic vapor to an ensemble of different two-level systems (accounting for the Zeeman transitions) of resonant frequency $\omega_i = \omega_e - \omega_g$. Each Zeeman transition gives rise to a resonance peak of amplitude proportional to $|\bra{e}d\ket{g}|^2$ and, thus, to $A_{eg}$.
Using this model allows to compute the intensity of any Zeeman transition of the $D_1$ and $D_2$ lines of any alkali atom and the absorption spectra of the vapor for arbitrary values of the external magnetic field. This model is applied in the following sections of this paper.

\section{Experiment}
\subsection{Atomic transitions under study: ${}^{87}$Rb $D_2$ line}

\begin{figure}
\centering
\input{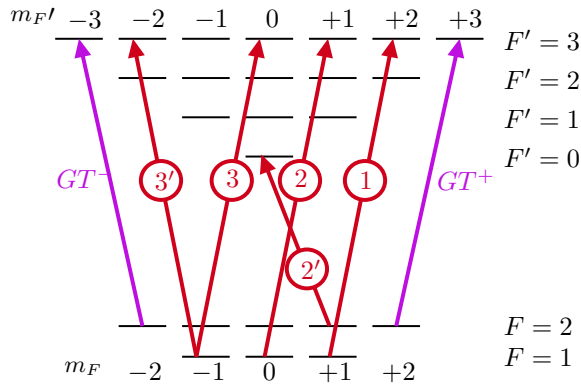}
\caption{Diagram of atomic transitions of $^{87}$Rb $D_2$ line. The $1\rightarrow 3'$ $\sigma^+$ MI transitions are numbered $1$-$3$, the strongest of them is marked with number $3$, transitions $\ket{2,2}\rightarrow \ket{3',3'}$  and $\ket{2,-2}\rightarrow \ket{3',-3'}$ are respectively labelled GT$^+$ and GT$^-$. The strongest $\sigma^-$ MI transitions are numbered $2'$ and $3'$.}
\label{fig:1}
\end{figure}

Figure \ref{fig:1} shows a diagram of the atomic transitions of the $D_2$ line of $^{87}$Rb. The group of MI transitions $1 \rightarrow 3'$ (hereafter, we consider only MI2 transitions, so we omit the number “2”) excited by $\sigma^+$ radiation are numbered $1$-$3$ in circles. The strongest of them, the transition $\ket{1,-1}\rightarrow \ket{3',0'}$ is marked with number $3$ in a circle. The $\sigma^+$ transition $\ket{2,+2}\rightarrow \ket{3',+3'}$  is labeled as GT$^+$ and the $\sigma^-$-transition $\ket{2,-2}\rightarrow \ket{3',-3'}$ is labeled as GT$^-$.

GT$^+$ and GT$^-$ transitions are called "guiding transitions": their intensities are constantly equal to each other and do not depend of the $B$-field since they occur between states that are not mixed by the magnetic field \cite{Sargsyan2015EPL}. These features of the GT transitions are used in the experiment described below to compare the amplitudes of MI transitions upon excitation by circularly polarized laser radiation ($\sigma^+$ or $\sigma^-$). Registering the spectra upon excitation by $\sigma^-$ (which contains MI transitions on the low-frequency wing of the spectrum) and by $\sigma^+$ radiation (which contains MI transitions on the high-frequency wing of the spectrum) and direct comparison of their amplitudes can lead to errors (the parameters of the used diode laser can change when scanning the frequency in a wide frequency range). Therefore, we compare the amplitude of the MI transition to the GT transition located close to it in frequency. Since the intensities of GT transitions for $\sigma^+$ or $\sigma^-$ radiations are equal to each other $A($GT$^+) = A ($GT$^-)$, this allows us to determine the intensities of the MI transitions of $^{87}$Rb of interest. This technique with the involvement of GT transitions was successfully used to study MI transitions of $^{85}$Rb and Cs in \cite{Sargsyan2021JETPlett,Sargsyan2021PLA}.

\subsection{Experimental setup}

\begin{figure}
\centering
\includegraphics[width=0.45\textwidth]{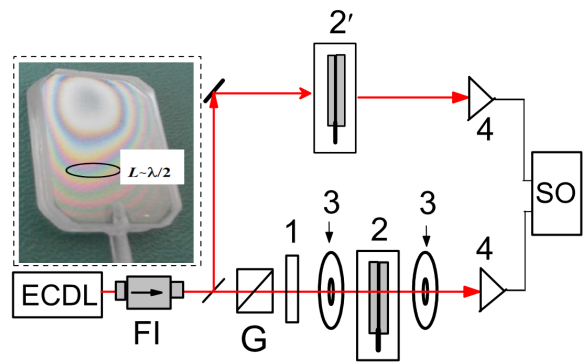}
\caption{Layout of the experiment: (ECDL) cw diode laser with the wavelength $\lambda =780$ nm, (FI) Faraday insulator, (1) quarter-wave plate, (2) nanocell filled with Rb, (3) permanent magnets, (4) photodetectors, (2') is an additional Rb NC (denoted NC2) with a thickness $L = 390$ nm, the SD absorption spectrum of which is a frequency reference spectrum, SO- digital oscilloscope, G- Glan polarizer. The inset shows the photograph of the Rb NC, the interference fringes formed at the reflection of light from inner surfaces of the windows are seen, oval marks a region with thickness $L~=\lambda /2 = 390$ nm.}
\label{fig:2}
\end{figure}

Figure \ref{fig:2} shows the layout of the experimental setup. A MOGLabs Cateye external-cavity diode laser (ECDL) with a wavelength of $780$ nm and a spectral width of around $100$ kHz was used. The laser beam diameter is 1 mm. To detect the transmission (absorption) spectrum, we used a nanocell (NC) filled with Rb atomic vapor of thickness $\lambda /2$ ($\simeq 390$ nm) along the direction of the laser radiation, $\lambda$ being the resonant wavelength of Rb $D_2$ line.

The NC was used to implement the $\lambda /2$ method, which ensured the narrowing of atomic transitions (lines) in the absorption spectrum $A(\nu)$ of the NC \cite{Dutier2002,Sargsyan2014LPL,Vartanyan1995}. To further narrow the atomic lines, we performed second derivative (SD) of the absorption spectrum $A''(\nu)$ \cite{Savitzky1964,Talsky1994,Sargsyan2019OL}. This is particularly important for the frequency separation of transitions in case some of them are overlapped. The NC was placed in a furnace with a hole allowing for the laser radiation passage and was heated to $120$~$^\circ C$ to ensure an atomic density $N \approx 2\times 10^{13}$ cm$^{-3}$ (details of the design of the nanocell are presented in \cite{Sargsyan2021PLA}). The main nanocell was placed between strong permanent magnets which produce a strong longitudinal magnetic field, and the wave vector of laser radiation $k$ was directed along the magnetic field $B$ \cite{Sargsyan2014}. 
To form a frequency reference, we used the SD of the absorption spectrum of an additional $390$ nm-thick cell (NC2) containing Rb, towards which part of the laser radiation was directed \cite{Sargsyan2019OL}. The absorption signals were recorded by FD-24K photodiodes (4), the signals from which were fed to a Tektronix TDS2014B (SO) oscilloscope.

\subsection{Experimental results and discussions}

\begin{figure}
\centering
\includegraphics[width=0.45\textwidth]{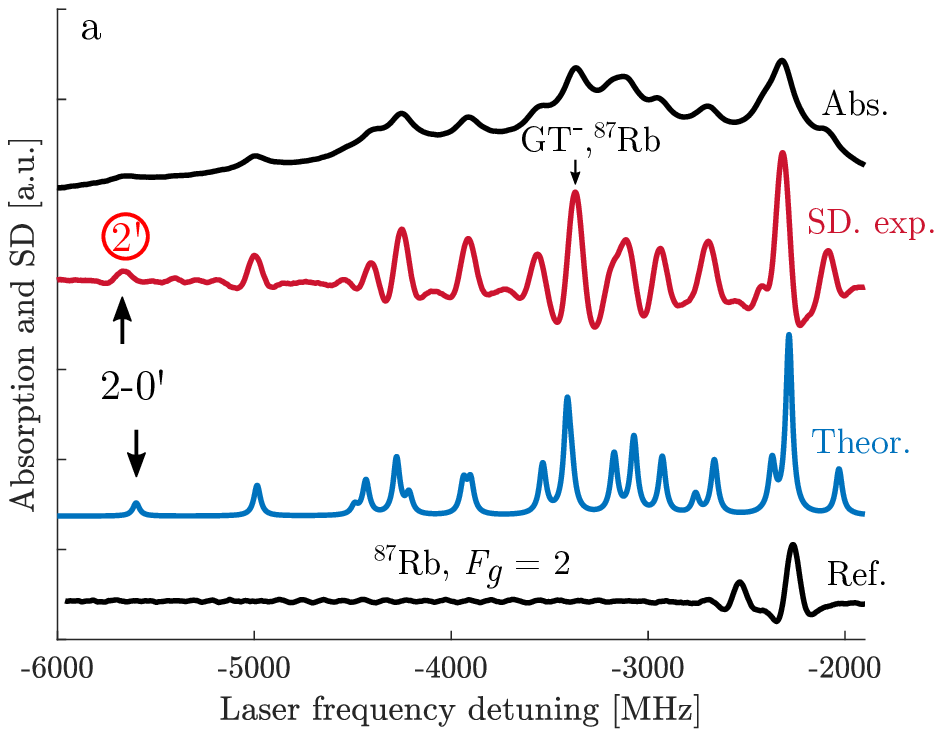}
\includegraphics[width=0.45\textwidth]{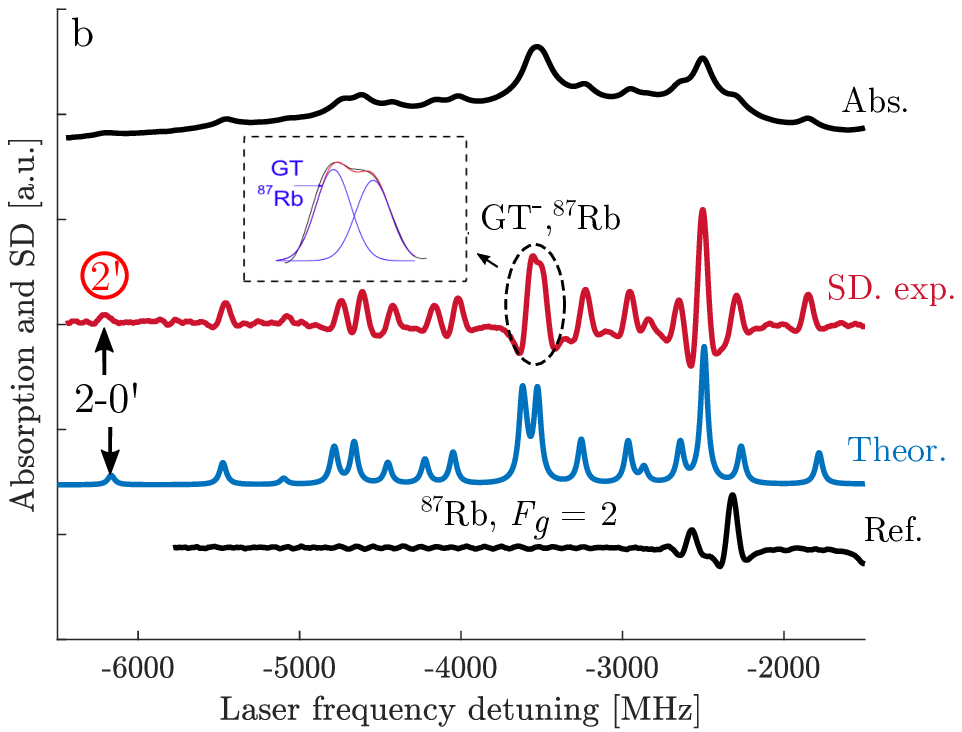}
\includegraphics[width=0.45\textwidth]{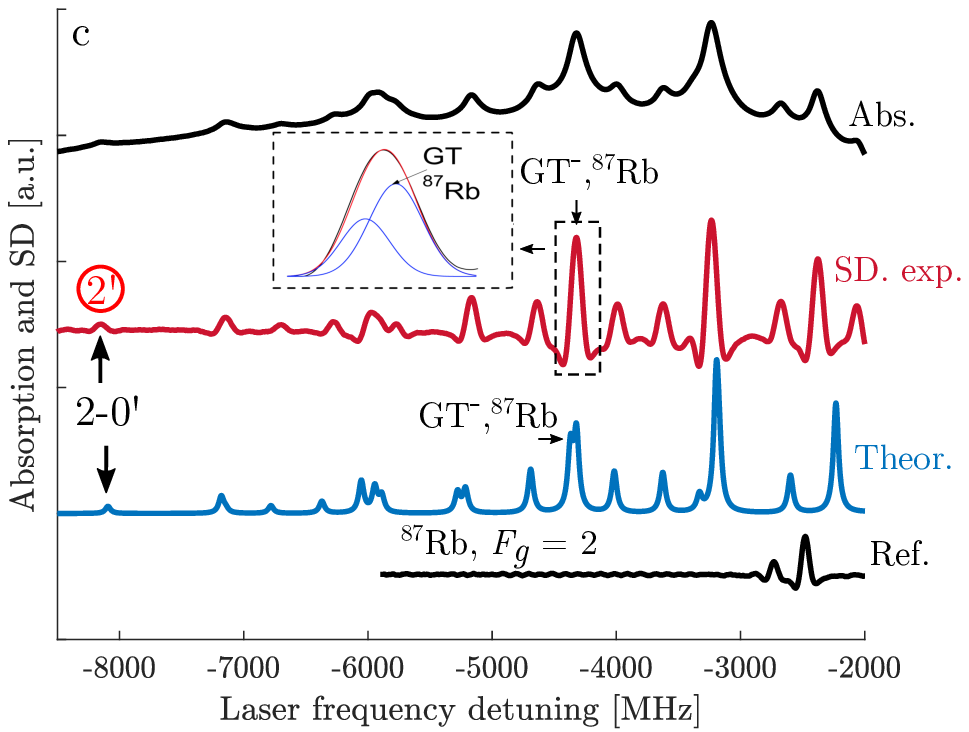}

   % \caption{$^{87}$Rb atom, $D_2$ line, $\sigma^-$ -excitation; NC thickness $L = 390$ nm; a) b) c) upper black lines are the absorption spectra, transitions $1,2\rightarrow 0',1',2',3'$ , $B = 850, 950$ and $1450$ G, respectively, red lines are SD spectra of the absorption, red digit 2 in circle marks the strongest of the indicated MI transitions GT$^-$ ($^{87}$Rb) is the guiding transition, blue lines are calculated SD absorption spectra ; a) b) c) (Ref)— lower lines are the SD absorption spectra of $^{87}$Rb, transitions $2\rightarrow 1', 2',3'$ , $B=0$.}

    \caption{Absorption spectra of transitions $1,2 \rightarrow 0',1',2',3'$ of the $D_2$ line of ${}^{87}$Rb for $\sigma^-$-polarized incident laser radiation. a) $B = 850$ G, b) $B = 950$ G, c) $B=1450$ G. The spectra were obtained with a cell of thickness $\simeq 390$ nm. For each figure, the upper black curve (Abs.) is the experimental absorption spectrum. The red curve is the SD of the absorption spectrum (SD. exp.), and the blue curve (Theor.) is theoretical. Lower black curve (Ref.) is the SD of the peaks corresponding to the transitions $2\rightarrow 1',2',3'$. The labelling GT$^-$ and $2'$ is consistent with Figure $\ref{fig:1}$. Some non-linearity of the laser frequency scanning is caused by the imperfect grating control of the laser.}

\label{fig:3}
\end{figure}

\begin{figure}
\centering
\includegraphics[width=0.45\textwidth]{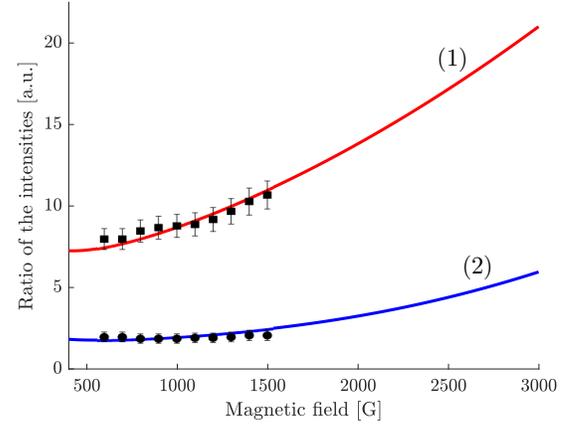}
\caption{Curve 1- ratio of the amplitude of GT$^+$  to the MI transition numbered $2'$ with respect to the magnetic field $B$. The theory and the experiment are in good agreement. Curve 2- ratio of the amplitude of GT$^+$ and MI transition numbered 3 with respect to the magnetic field $B$. The theory and the experiment are in good agreement.}
\label{fig:5}
\end{figure}

The upper black curves in Fig. \ref{fig:3} a,b,c (Abs.) are the experimental absorption spectra of transition $1,2\rightarrow 0',1',2',3'$ of the $D_2$ line of $^{87}$Rb for the following magnetic field values: $B =850$ G (a), $B =950$ G (b), and $B=1450$ G (c) obtained by the $L =\lambda /2 = 390$ nm method when $\sigma^-$ polarized radiation is applied (transitions frequencies are shifted towards to low frequencies as the magnetic field increases, this behavior has been throughly studied for sodium in \cite{MomierJQSRT2021}). The laser power was $30$ $\mu$W. As can be seen, some transitions in the absorption spectra are partially overlapped. Red lines are SDs of absorption spectra (here and below, the SD is inverted for convenience). The transition $\ket{2,+1}\rightarrow \ket{0',0'}$ numbered $2'$ is among the strongest MI transitions (together with transition numbered $3'$ shown in Fig. \ref{fig:1}). The spectrum also includes the guiding transition GT$^-$ ($^{87}$Rb) whose application is discussed above. We need to determine the ratio of the amplitude of the GT$^-$ transition to the amplitude of the transition numbered $2'$, in the cases indicated in Fig. \ref{fig:3}b and \ref{fig:3}c, the GT$^-$ transition is overlapped with other atomic transitions; therefore, the insets show the fitted spectra. The ratios of the amplitude of the GT$^-$ transition to the amplitude of the transition numbered $2'$ for $850$, $950$, and $1450$ G (experimental results are also given for other values of $B$) are presented in the upper curve (1) in Fig. \ref{fig:5}. Blue lines are the SDs of theoretical absorption spectra for atomic transitions with a FWHM (full width at half maximum) of $40$ MHz. As is seen, there is a good agreement between experiment and theory regarding the amplitude and the (frequency) position of the peaks.

\begin{figure}
\centering
\includegraphics[width=0.45\textwidth]{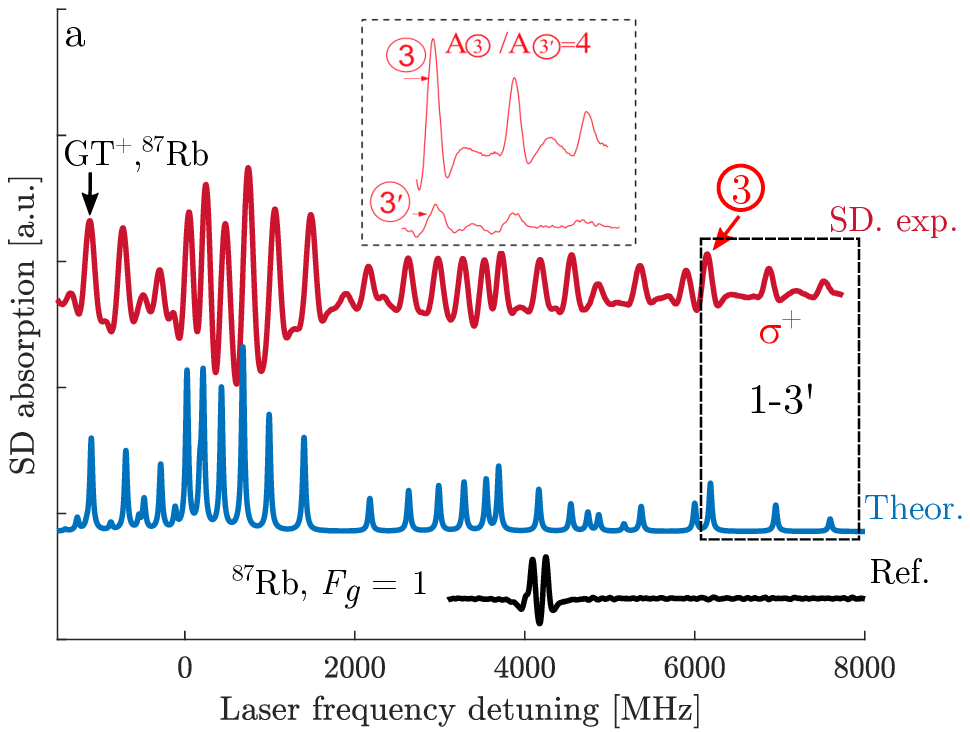}
\includegraphics[width=0.45\textwidth]{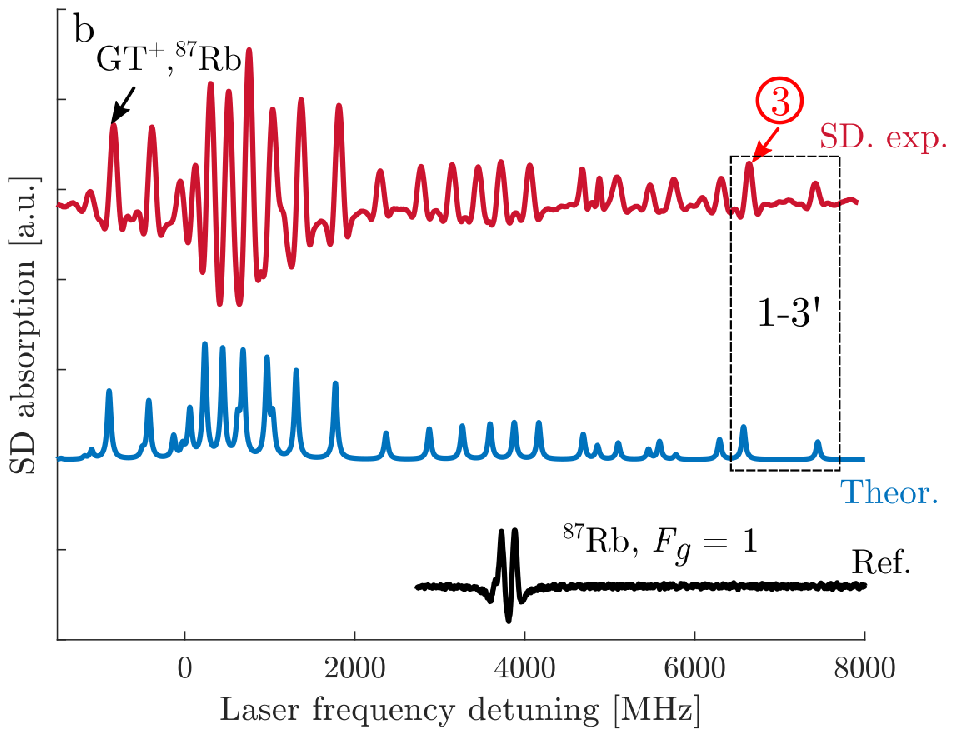}
\includegraphics[width=0.45\textwidth]{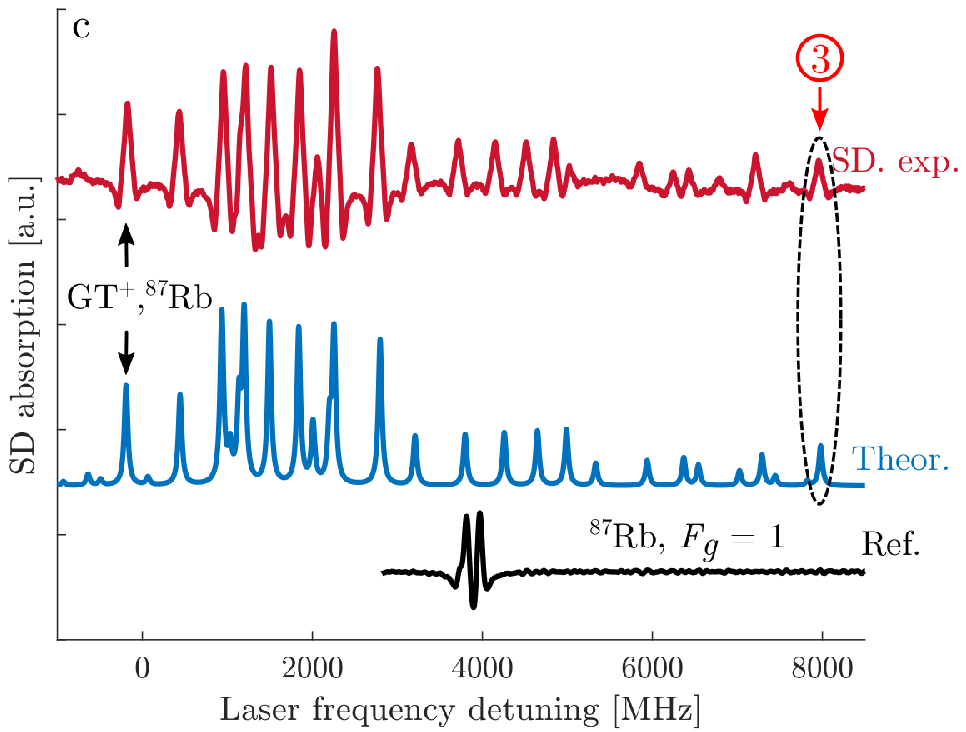}

%    \caption{$^{87}$Rb $D_2$ line spectra for $\sigma^+$-excitation obtained with a cell of thickness $L = 390$ nm; a) b) c) - red lines are SD the absorption spectra, transitions $1,2 \rightarrow 0',1', 2', 3'$; $B =$ (a) $850$, (b) $1000$, and (c) $1500$ G, MI transition in digit 3 in circle marks the strongest MI transition of the group transitions $1\rightarrow 3'$; GT$^+$ ($^{87}$Rb) - guiding transition; blue lines are calculated SD absorption spectra; a) b) c) lower lines (Ref.) are the SDs of the absorption spectra $1\rightarrow 0', 1',2'$ transitions in $^{87}$Rb, $B=0$, inset in (a) shows the SD spectrum of the MI transition numbered 3 in red circle for $\sigma^+$- excitation and MI transition numbered $3'$ in red circle for $\sigma^-$- excitation (both are normalized using GT$^+$ and GT$^-$ transitions), the ratio of the amplitudes is equal to 4.}
    \caption{SD of absorption spectra of transitions $1,2 \rightarrow 0',1',2',3'$ of the $D_2$ line of ${}^{87}$Rb for $\sigma^+$-polarized incident laser radiation. a) $B = 850$ G, b) $B = 950$ G, c) $B=1450$ G. The spectra were obtained with a cell of thickness $\simeq 390$ nm. For each figure, the red curve is the SD of the experimental absorption spectrum (SD. exp.), and the blue curve (Theor.) is theoretical. Lower black curve (Ref.) is the SD of the peaks corresponding to the transitions $1\rightarrow 0',1',2'$. The labelling GT$^+$ and $3$ is consistent with Figure $\ref{fig:1}$. The inset in a) shows the SD of transitions $3$ and $3'$ (both are normalized to the amplitudes of GT$^+$ and GT$^-$): the ratio of amplitudes is equal to 4.}
\label{fig:4}
\end{figure}

The calculations were performed using the theoretical models presented before and in \cite{Tremblay1990,Sargsyan2020JPB,Sargsyan2014,Sargsyan2017,Dutier2002}. Lower lines (Ref.) in Fig. \ref{fig:3} show the SD of the absorption spectra $2 \rightarrow 1',2',3'$ transitions in $^{87}$Rb for $B=0$. 

The upper red lines in Fig. \ref{fig:4} a,b,c are SD of the experimental absorption spectra (not shown) for the $\sigma^+$ transitions $1, 2 \rightarrow 0',1', 2', 3'$ , at $B =850$ G (a), $B = 1000$ G (b) and $1500$ G (c) obtained by the $L =  \lambda /2 = 390$ nm method (transitions frequencies are shifted from the zero field frequencies towards the high-frequency wing of the spectrum \cite{MomierJQSRT2021}). The transition $\ket{1,-1}\rightarrow\ket{3',0'}$ labelled 3 in red marks the strongest $\sigma^+$ MI transition. The spectra also include the guiding transition GT$^+$. We need to determine its amplitude but there is no need to fit the absorption peak corresponding to this transition since it isn't overlapped with any other transition. The ratio of the amplitude of the GT$^+$ transition to the amplitude of the transition labelled $3$ for $850$, $1000$, and $1500$ G are presented in the curve 2 of Fig. \ref{fig:5}  (experimental points for other values of $B$ are also provided). Blue lines in Fig. \ref{fig:4} a,b,c are the SD of the calculated absorption spectra with a FWHM of $40$ MHz. It can be seen here again that good agreement between the theory and experiments is obtained. Lower lines (Ref.) in Fig. \ref{fig:4} a,b,c are the SD of the absorption spectra of transitions $1 \rightarrow 0', 1',2'$ in zero magnetic field. The inset in Fig. \ref{fig:4}a shows the SD spectrum of the MI transition numbered 3 for $\sigma^+$ polarized radiation and MI transition numbered $3'$ for $\sigma^-$ polarized radiation (which are normalized to the amplitudes of GT$^-$ and GT$^+$), the ratio of the amplitudes is equal to 4. The frequency distance between them is around $3$ GHz for $B \approx 1000$ G. Nevertheless, for comparison, we brought them together.

\begin{figure}
\centering
\includegraphics[width=0.45\textwidth]{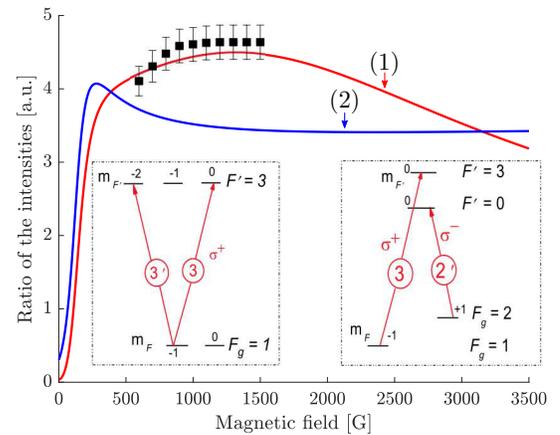}
\caption{Curve 1- ratio of the amplitudes of MI transition numbered $3$ in red circle to the transition numbered $2'$ with respect to the magnetic field $B$. The theory and the experiment are in good agreement. Curve 2 -calculated ratio of the amplitudes of transition numbered $3$ to transition numbered $3'$ with respect to the magnetic field $B$.}
\label{fig:6}
\end{figure}

As it is seen from the inset, the ratio of the amplitudes of MI transitions $A(3)/A(3')$ is equal to $4$, which coincides with the value predicted theoretically and measured experimentally \cite{Tonoyan2018EPL}. In Fig. \ref{fig:5} curve 1 is the ratio of amplitudes of the GT$^+$ transition to the MI transition numbered $2'$  with respect to the magnetic field $B$, the theory and the experimental are in perfect agreement. The ratio of the amplitude of the GT$^+$ transition to the amplitude of the MI transition numbered $3$ for $B =850$, $1000$, and $1500$ G is presented by the curve 2 in Fig. \ref{fig:5} (experimental results are also given for other values of $B$). The ratio of the amplitude of the MI transition numbered 3 to the amplitude of the transition numbered $2'$ with respect to the magnetic field $B$ is presented by the curve 1 in Fig. \ref{fig:6} (both the theory and the experiment). Calculated ratio of the amplitude of the transition numbered $3$ to the amplitude of the transition numbered $3'$ as a function of the magnetic field is presented by the curve 2 in Fig. \ref{fig:6}. As seen from curve 1 in Fig. \ref{fig:6} for $B >500$ G, the probability of the strongest $\sigma^+$ MI transition ($F_g =1 \rightarrow F_e =3$) is 4.5 times higher than the probability of the strongest $\sigma^-$ MI transition.

\section{Conclusion}

The probabilities of atomic transitions $F_e - F_g = \Delta F = \pm 2$ between a ground $F_g$ and an excited $F_e$ level of the hyperfine structure of the $D_2$ line of alkali metals are zero when no magnetic field is applied. In the range of $0.1$ - $3$ kG, there is a gigantic increase in their probabilities, therefore, they are called Magnetically Induced (MI) transitions. It is demonstrated (both experimentally and theoretically) that for an atom, particularly for $^{87}$Rb atoms with a nuclear spin $I = 3/2$ (same for $^{39}$K, $^{23}$Na, $^7$Li) in magnetic fields larger than $100$ G, intensity of the strongest $\sigma^+$ MI transition $1 \rightarrow 3'$ is $4.5$ and $4$ times higher than the probability of the strongest $\sigma^-$ MI transition $2 \rightarrow 0'$ and $1 \rightarrow 3'$ respectively. We call this difference is as Type 2 Magnetically Induced Circular Dichroism, which is stronger expressed than that of the Cs and $^{85}$Rb atoms $D_2$ lines.
Thus, it is important to note that the $\sigma^+$ MI transition $1\rightarrow 3'$ is very promising for applications of magneto-optical processes occuring in strong magnetic fields.
\acknowledgments
\noindent This work was supported by the Science Committee of RA, in the frame of the research project n°~21T-1C005.

\bibliography{biblio}% Produces the bibliography via BibTeX.

\end{document}